\documentclass[twocolumn]{article}  
\usepackage[a4paper,margin=1in]{geometry}
\date{}

\usepackage{amsmath,amssymb,mathtools,bm}
\usepackage{amsthm}
\usepackage{mathrsfs}
\usepackage{stackrel}
\usepackage{nicefrac}
\usepackage[normalem]{ulem}

\usepackage{physics}

 \theoremstyle{remark}	

  \theoremstyle{remark}	
 
 \theoremstyle{remark}	

 \theoremstyle{remark}	

 \theoremstyle{remark} 

 \theoremstyle{remark} 

 \theoremstyle{remark} 

\usepackage[normalem]{ulem}
\usepackage[utf8]{inputenc}  
\usepackage{bbm}
\usepackage{etoolbox}

\usepackage{comment}

\usepackage{hyperref}
\hypersetup{colorlinks=true,linkcolor=blue,citecolor=blue,urlcolor=blue}

\usepackage{graphicx}
\usepackage{tikz}
\usetikzlibrary{arrows.meta, positioning}

\usepackage{authblk}
\usepackage{orcidlink} 

\author[1,2,\orcidlink{0009-0008-5109-1922}]{Amit Kam}
\author[3,4,\orcidlink{0000-0002-1168-5944}]{Charles Roques-Carmes}
\author[5,\orcidlink{0000-0003-0167-3402}]{Shai Tsesses}
\author[2,6,\orcidlink{0000-0002-4056-4455},*]{and Aviv Karnieli}

\affil[1]{Department of Physics, Technion - Israel Institute of Technology, Haifa 32000, Israel}
\affil[2]{Helen Diller Quantum Center, Technion - Israel Institute of Technology, Haifa 32000, Israel}
\affil[3]{E. L. Ginzton Laboratories, Stanford University, 348 Via Pueblo, Stanford, CA USA}
\affil[4]{Institute of Science and Technology Austria (ISTA), Klosterneuburg 3400, Austria}
\affil[5]{Department of Physics and Research Laboratory of Electronics, Massachusetts Institute of Technology, Cambridge, Massachusetts 02139, USA}
\affil[6]{Andrew and Erna Viterbi Department of Electrical \& Computer Engineering, Technion - Israel Institute of Technology, Haifa 32000, Israel}
\affil[*]{Corresponding author email address: 
\href{mailto:karnieli@technion.ac.il}{karnieli@technion.ac.il}}

\usepackage[
    backend=biber,
    style=chem-acs,
    sorting=none,
    articletitle=true,
    doi=true,
    url=false,
    isbn=false
]{biblatex}

\let\cite\supercite

\usepackage{totcount}
\newtotcounter{citedrefs}

\AtEveryBibitem{\stepcounter{citedrefs}}

\defbibnote{referencecount}{%
    \noindent\textit{This article references \total{citedrefs} other publications.}
    \par\medskip
}

\title{Quantum Skyrmions in Mixed States of Light \\ and their Nested Topology}

\begin{document}
\twocolumn[
\maketitle
\begin{center}
  \begin{minipage}{0.88\textwidth}
    \small  
    \begin{abstract}
    Topological quasiparticles of light, such as classical and quantum optical skyrmions, have so far relied on fully coherent or pure quantum states whose topology is encoded in the entanglement between polarization and two-dimensional spatial modes.
    Here we show that skyrmionic topology can emerge directly within the density matrix of a mixed quantum state. We introduce a framework in which a coherence–Stokes vector defines a topological texture over the density matrix, enabling the realization of quantum skyrmions using only a pseudospin and a real or synthetic one-dimensional space of modes. For a single photon, the density matrix is analogous to the coherence matrix of classical light, and can be encoded using partially coherent electromagnetic fields. We further analyze the topological texture encoded in a bipartite entangled photon pair, showing how skyrmions arise not only in the full bi-photon system, but also in its reduced subspaces with any pseudospin-mode combination simultaneously. We then explore the robustness of such skyrmions to environmental noise, and discover their persistence in mixed-quantum states of multiple photons. Finally, we propose a feasible experimental route to generate and measure such skyrmions using integrated photonic networks, and suggest avenues for similar implementations in other quantum systems. Our work paves the way for robustly encoding classical information on partially coherent light or on mixed quantum states and for encoding topological charges on many-body quantum systems.
    \end{abstract}
  \end{minipage}
\end{center}
]

\section*{Keywords}
Quantum optical skyrmions; Partially coherent light; Mixed quantum states; Quantum coherence; Quantum state engineering; Multipartite entanglement.

\section*{Introduction}
Skyrmions are topologically protected quasiparticles with wide-ranging relevance across fields~\cite{skyrme1961non}, from particle~\cite{skyrme1962unified} and condensed-matter physics~\cite{fert2017magnetic} to photonics~\cite{shen2024optical}. In particular, optical skyrmions were first observed in evanescent electromagnetic fields~\cite{tsesses2018optical,du2019deep,davis2020ultrafast}, and since then have been realized across  photonic platforms in a variety of degrees of freedom, such as Stokes-parameter vectors (polarization)~\cite{gao2020paraxial,sugic2021particle,shen2022generation,spaegele2023topologically,wei2026nanophotonic,liu2026broadband}, magnetic fields~\cite{deng2022observation}, energy-flow fields~\cite{shen2021supertoroidal}, and other pseudospin vectors~\cite{karnieli2021emulating, wang2022topological}.

Recent works have highlighted the relevance of optical skyrmions to quantum information science~\cite{ornelas2024non, ma2025nanophotonic, kam2026quantum,de2025revealing,ma2026high}, encoding them on single-photon or entangled biphoton states via structured spatial modes of light, and demonstrating that the encoding remains robust under typical noise channels, which smoothly deform the underlying wavefunction~\cite{wang2024topological,guo2026topological,ornelas2025topological,de2025quantum}.

\begin{figure*}[ht]
    \centering
    \includegraphics[width=\textwidth]{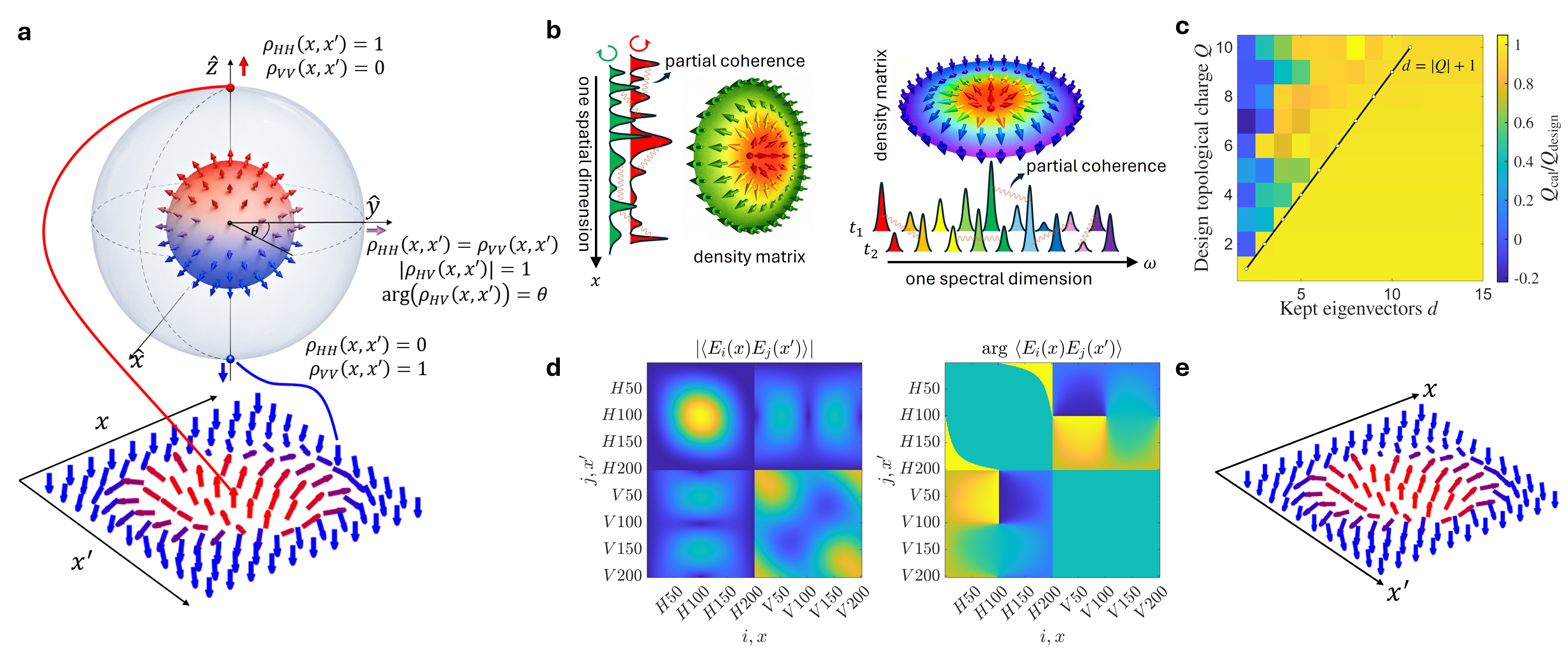}
        \caption{\textbf{Skyrmions encoded in the partial coherence of light.} \textbf{(a)} Schematic of a skyrmion encoded in the density matrix of one-dimensional mixed state of light. The two density-matrix coordinates $(x,x')$ form an effective two-dimensional parameter space, over which the normalized coherence-Stokes vector $\mathbf{s}(x,x')$ defines a skyrmionic texture and maps onto the Poincaré sphere. \textbf{(b)} An optical skyrmion encoded in one-dimensional partially coherent light, in either real space or synthetic dimension (such as frequency). \textbf{(c)} The designed skyrmion number is shown as a function of the number of retained eigenvectors in the density matrix. In particular, retaining $(d=|Q|+1)$ eigenvectors is found numerically to be sufficient to faithfully reproduce the target skyrmion. \textbf{(d)} A density matrix is constructed from the analytic solution with eigenmodes defined in Eqs.~\ref{eq:u_modes}. \textbf{(e)} From this density matrix, we compute the coherence Stokes vector and plot its associated skyrmion texture for $M=11$ spatial modes. The choice of $M=11$ is intended as a compact visualization of the texture, and is sufficient for the calculation of the skyrmion number via the Berg-Lüscher \cite{berg1981definition} method. A higher number of modes is needed for high-precision convergence of the differential integral, Eq.~\ref{eq:skyrmion_number} (see also Supplementary Material section S3).}
    \label{fig:1}
\end{figure*}

Common to all previous realizations is the use of fully coherent and polarized light: either pure quantum states or coherent (classical) states, yet many practical sources are prone to decoherence, resulting in probabilistically-mixed states. Such partially coherent light can still be useful, and has recently shown promise for applications such as neuromorphic computation~\cite{dong2024partial} and holography~\cite{lee2020light,pang2025coherence}. This naturally raises the question of whether topological quasiparticles of light can be realized with partially coherent light, and, if so, whether they persist in noisy quantum channels, where entanglement usually breaks down.

Here, we propose a scheme for encoding skyrmions in the density matrix describing a fully quantum system and theoretically explore their unique topological properties. For a single quantum particle, the density matrix is analogous to the coherence matrix of classical light, through which we show that skyrmions can be encoded using partially coherent electromagnetic fields described by a pseudospin and a one-dimensional space of modes in either real (e.g., spatial-bin) or a synthetic (e.g., frequency-bin) dimension. Generalizing this result further, we analyze the topological texture encoded in a bipartite entangled state, such as an entangled photon pair, demonstrating that skyrmions exist in the full system, and in its reduced subspaces with any pseudospin-mode combination (both local and non-local). We then explore the  robustness of such skyrmions to noise and to loss of entanglement, and further discover bipartite skyrmions in mixed-quantum states of a many-body system, which can be constructed by multiple photons. Finally, we propose a feasible experimental route to generate and measure such skyrmions using integrated photonic networks, and suggest avenues for similar implementations in other quantum systems, such as trapped ions or neutral atoms. Our work paves the way for robustly encoding classical information on partially coherent light or on mixed quantum states and for imprinting topological charges on many-body quantum states.

\section*{Results}

We use standard quantum notation conventions as summarized in the Supplementary Material section S1.

\textbf{\textit{Encoding a skyrmion on a single-photon density matrix.}} We start by considering the encoding scheme of a skyrmion texture on the density matrix $\hat{\rho}$ of a single photon, which is equivalent to the first-order coherence matrix of partially-coherent classical light. To this end, we consider a field with an ``internal'' pseudospin degree of freedom (e.g., polarization, time-bin, etc.), and a coordinate $x$ - which can be realized, for example, by encoding $M$ modes in either real space (e.g., modes in a 1D waveguide array), or in a synthetic dimension (e.g., multiple frequency bins), as illustrated in Fig.~\ref{fig:1}(b). Unlike previous implementations of coherent optical skyrmions, encoding the texture in the space created by two physical coordinates $x$ and $y$, requiring $M$ modes per coordinate, or $M^2$ modes in total, our proposal for partially-coherent skyrmions is realized using only a single physical coordinate $x$, and requires only $M$ modes. This enables to reduce the physical dimensionality for encoding a skyrmion texture by using fewer modes, without losing the topological structure.

For simplicity, we shall henceforth consider the internal degree of freedom to be polarization, and denote it by $\sigma \in \{H,V\}$, and the coordinate $x$ to comprise a set of $M$ spatial bins. Motivated by the Stokes-parameter vector construction of optical skyrmions~\cite{gao2020paraxial,sugic2021particle,ornelas2024non}, we define Stokes-like components from the single-photon density matrix $\hat{\rho}$ via
\begin{equation}
    \begin{split}
    &S_x(x,x')+iS_y(x,x') = \rho_{HV}(x,x'),\\
    & S_z(x,x') = |\rho_{HH}(x,x')| - |\rho_{VV}(x,x')|,
    \end{split}
    \label{eq2}
\end{equation}

where $\rho_{\sigma\sigma'}(x,x')=\langle 1_{\sigma x}|\hat{\rho}|1_{\sigma' x'}\rangle$ denotes the density matrix elements in the single photon Fock-state basis. As illustrated in Fig. 1a, the mapping between the density matrix and the Stokes vector $\mathbf{S}(x,x')=(S_x ,S_y ,S_z)$ could be understood geometrically as follows. For a point $(x,x')$ of the density matrix, if the population of the state is fully in $H$ or $V$, $\rho_{HH}(x,x')=1,~\rho_{VV}(x,x')=0$ ($\rho_{HH}(x,x')=0,~\rho_{VV}(x,x')=1$), then the Stokes vector points to the north (south) pole of the Poincare sphere. If the populations are equal $\rho_{HH}(x,x')=\rho_{VV}(x,x')$, the Stokes vector points to the equator of the Poincare sphere, with the azimuthal direction determined by the argument of the coherence $\rho_{HV}(x,x')$.

We note that in general, a density matrix $\rho_{\sigma\sigma'}(x,x')$ will not produce a normalized Stokes vector satisfying $|\mathbf{S}(x,x')|=1$. We normalize $\mathbf{S}$ by defining the local magnitude
\begin{equation}
S_0(x,x')=\sqrt{S_x^2(x,x')+S_y^2(x,x')+S_z^2(x,x')}.
\label{eq:2}
\end{equation}
The vector $\mathbf{S}(x,x')$ is a valid texture if the normalization $S_0$ remains nonzero for all $(x,x')$. This requires that $\rho_{\sigma\sigma'}(x,x')$ is never a maximally unpolarized state, for which $\rho_{HH}(x,x')=\rho_{VV}(x,x'),~\rho_{HV}(x,x')=0$. The resulting normalized vector field
\begin{equation}
\mathbf{s}(x,x')=\frac{\mathbf{S}(x,x')}{S_0(x,x')}.
\end{equation}
supports an integer-valued skyrmion topological charge,
\begin{equation}
Q=\frac{1}{4\pi}\int \mathbf{s}\cdot
\left(\frac{\partial \mathbf{s}}{\partial x}\times
\frac{\partial \mathbf{s}}{\partial x'}\right)\,dx\,dx'.
\label{eq:skyrmion_number}
\end{equation}

A $2M\times 2M$ density matrix $\rho_{\sigma\sigma'}(x,x')$ is positive semidefinite (PSD) and, in principle, admits a spectral decomposition into $2M$ coherent eigenmodes~\cite{gamo1964iii,wolf1982new,salem2004polarization,gbur2002spreading}. Remarkably, realizing a skyrmion texture requires a much sparser representation. 

One can find a sparse representation using the following procedure. An exact skyrmion texture is constructed from an auxiliary Hermitian matrix $A_{\sigma \sigma'}(x,x')$. An example for such a matrix is
\begin{equation}
    \begin{split}
    &A_{HV}(x,x') = \sin\left[\Theta\left(\sqrt{x^2+x'^2}\right)\right]e^{i\Phi(x,x')},\\
    &A_{HH}(x,x') = \cos^2\left[\Theta\left(\sqrt{x^2+x'^2}\right)/2\right],\\
    &A_{VV}(x,x') = \sin^2\left[\Theta\left(\sqrt{x^2+x'^2}\right)/2\right].
    \end{split}
\end{equation}
where $\Theta(r)=\pi (r/R_0)\mathrm{H}(R_0-r)+\pi\mathrm{H}(r-R_0)$, $\Phi(x,x')= l\arctan(x,x')+\Phi_0$, $\mathrm{H}(x)$ denoting the Heaviside step function, $l$ is an integer winding number, $\Phi_0$ a constant phase, and $A_{\sigma\sigma'}(x,x')=A^*_{\sigma'\sigma}(x',x)$. Substituting $A_{\sigma \sigma'}(x,x')$ into Eq.~\ref{eq2} instead of $\rho_{\sigma\sigma'}(x,x')$ results in a desired skyrmion texture with charge $Q=-l$, of either Néel ($\Phi_0=0$) or Bloch ($\Phi_0=\pi/2$) types \cite{liu2016skyrmions}. Unfortunately, the auxiliary matrix $A_{\sigma \sigma'}(x,x')$ is not a valid density matrix, since it is not necessarily PSD.

To extract a valid PSD density matrix encoding the same topological texture, a spectral decomposition of $A_{\sigma \sigma'}(x,x')$ is performed and only the eigenvectors $u_{i\sigma}(x)$ with the first few largest positive eigenvalues $\lambda_i$ are kept. Finally, the density matrix is constructed from the kept eigenvectors of $A_{\sigma \sigma'}(x,x')$ as $\rho_{\sigma\sigma'}(x,x') = \sum_{i=1}^d \lambda_i u_{i\sigma}(x)u_{i\sigma'}^*(x')$. 

Here, $d$ is the number of kept positive eigenvalues such that the constructed texture in the density matrix has the designed topological charge $Q$. As can be seen from Fig.~\ref{fig:1}(c), we find numerically that choosing $d=|Q|+1$ guarantees the construction of the designed skyrmion. We further find numerically that for skyrmion numbers $|Q|\leq 5$ it is possible to replace the $d=|Q|+1$ eigenvalues $\lambda_i$ with equal weights $1/d$, which we shall do hereafter unless stated otherwise. We note that by construction, the normalization $S_0$ (Eq.~\ref{eq:2}) of the Stokes vector of the auxiliary matrix $A_{\sigma\sigma'}(x,x')$ satisfies $S_0(x,x')\equiv1$ for all $(x,x')$. For the constructed state $\rho_{\sigma \sigma'}(x,x')$, it can be verified numerically that the normalization remains nonzero for all $(x,x')$, thus defining a valid normalized texture $\mathbf{s}$.

\begin{figure*}[ht]
    \centering
    \includegraphics[width=2.2\columnwidth,  trim=1.5cm 0cm 0cm 0cm, clip]{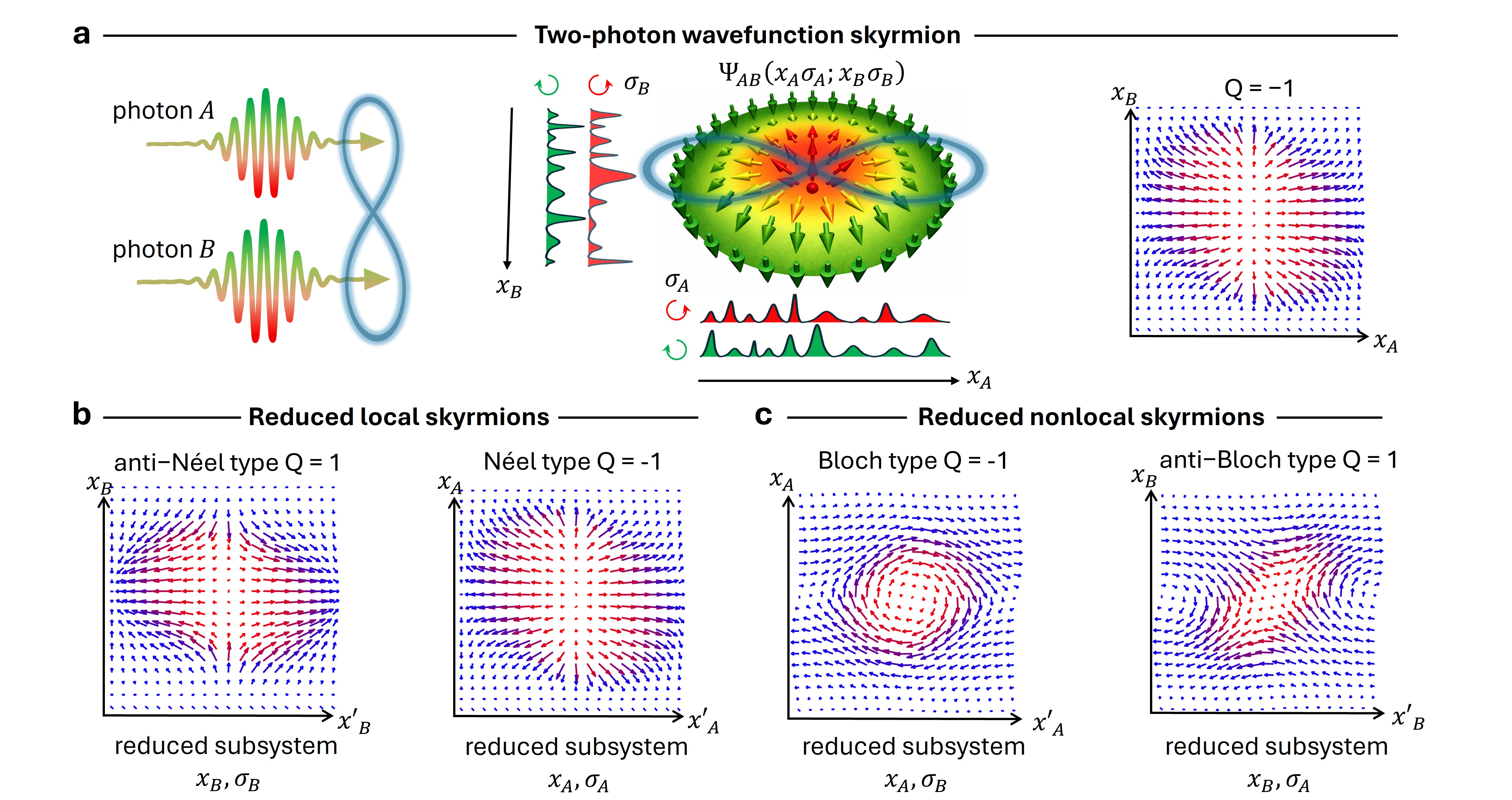}
        \caption{\textbf{Nested Topology of a two-photon entangled skyrmion encoded in the partial coherence of light.} \textbf{(a)} Two photons, A and B, each possess a polarization ${\sigma_i}$ and a spatial  degree of freedom $x_i$. The joint two-photon wavefunction, given in Eq.~\ref{eq:Psi_AB}, is engineered to carry a topological charge $Q=-1$, yielding the skyrmion texture shown in the illustration. \textbf{(b)-(c)} The system has four reduced subspaces, formed by tracing different pairs of polarization and spatial-mode sectors, each revealing an emergent skyrmion texture: \textbf{(b)} two local skyrmions appearing in the reduced density matrices of photons A and B individually, and \textbf{(c)} two non-local (entangled) skyrmions in the hybrid subspace made up of the polarization of one photon and the spatial mode of the other. Interestingly, each of these four subspaces generates a distinct skyrmion texture, corresponding to a different skyrmion type.}
    \label{fig:2}
\end{figure*}

As an example that admits an analytic solution, we consider the density matrix of a first-order skyrmion ($|Q|=1$), which can be constructed using only two mutually-incoherent orthogonal eigenstates, resulting in
\begin{equation}
\hat{\rho}=\frac{1}{2}\ket{u_{1}}\bra{u_{1}}
+\frac{1}{2}\ket{u_{2}}\bra{u_{2}}.
\label{eq:denisty matrix}
\end{equation}
where 
\begin{equation}
\ket{u_{i}}=\ket{u_{iH}}\ket{H}+\ket{u_{iV}}\ket{V}  
\label{u_dis}
\end{equation}
is the state-vector representation of $u_{i\sigma}(x)$. For a Néel skyrmion \cite{liu2016skyrmions} of topological charge $Q=-1$ or anti-Néel skyrmion with $Q=1$, the mode components are approximated analytically as
\begin{equation}
\begin{aligned}
\braket{x}{u_{1H}} &= \mp \frac{i}{2\sqrt{\mathcal{N}_1}}\sin\!\left(\pi \frac{x}{x_{\max}}\right),\\
\braket{x}{u_{1V}} &= \frac{1}{\sqrt{\mathcal{N}_1}}
\left[1+\sin^2\!\left(\frac{\pi}{2}\frac{x}{x_{\max}}\right)\right],\\
\braket{x}{u_{2H}} &= \frac{1}{\sqrt{\mathcal{N}_2}}
\left[1-\sin^2\!\left(\frac{\pi}{2}\frac{x}{x_{\max}}\right)\right],\\
\braket{x}{u_{2V}} &= \frac{1}{2\sqrt{\mathcal{N}_2}}\sin\!\left(\pi \frac{x}{x_{\max}}\right).
\end{aligned}
\label{eq:u_modes}
\end{equation}
for $|x|\leq x_{\mathrm{max}}$, and where the normalization constants $\mathcal{N}_i$ are set such that each eigenmode is unit-normalized $\braket{u_i}{u_i}=1$. An illustration of the density matrix with the resulting skyrmionic texture is given in Figures ~\ref{fig:1}(d)-(e). The above construction can be generalized to higher-order skyrmions and other topological quasiparticles of light. See Supplementary Material, Sections~S2 and~S2.1, for the realization of higher-order skyrmion textures, as well as the encoding of skyrmionium textures in coherence space.\\

We note that the accuracy of evaluating the skyrmion number depends on the number of modes $M$ used in the construction of the skyrmion, as they define the numerical resolution for the calculation via the differential method (Eq.~(\ref{eq:skyrmion_number})). In the Supplementary Material, section S4 we find numerically the minimal number of modes necessary for the calculation of an integer skyrmion number to within $2\%$ error. We do this for the differential method of Eq.~(\ref{eq:skyrmion_number}) as well as other numerical approaches for computing the skyrmion number that do not involve numerical differentiation, and require less modes: the Berg-Luscher method \cite{berg1981definition} and the McWilliam method \cite{mcwilliam2023topological}.\\

\textbf{\textit{Encoding a skyrmion on a two-photon wavefunction.}} The encoding scheme outlined above could be extended to a two-photon state with a form depending on the topological charge $|Q|$. For this we consider the maximally entangled state
\begin{equation}
\ket{\Psi}_{A,B}=\sum_{i=1}^{d}\frac{1}{\sqrt{d}}\ket{u_i}_A\ket{u_{i}^{*}}_{B} .
    \label{psi}
\end{equation}
The two-photon wavefunction also admits a skyrmionic distribution with respect to the bipartite system coordinates $\sigma_A,x_A$ and $\sigma_B,x_B$. For example, in the case of $Q=\pm1$ the wavefunction becomes 
\begin{equation}
\begin{split}
    \Psi_{AB}(\sigma_A,x_A;\sigma_B,x_B)= \frac{1}{\sqrt{2}}\,u_{1\sigma_A}(x_A)\,u_{1\sigma_B}^{*}(x_B)\\
    +\frac{1}{\sqrt{2}}\,u_{2\sigma_A}(x_A)\,u_{2\sigma_B}^{*}(x_B),
\end{split}
\label{eq:Psi_AB}
\end{equation}
having the same structure as Eq. \ref{eq:denisty matrix}. This two-photon entangled skyrmion generalizes previous realizations using entangled states of spin and orbital angular momentum~\cite{ornelas2024non}, where the skyrmion was defined non-locally using the polarization degree of freedom in one photon and the spatial degree of freedom of the second photon~\cite{ma2025nanophotonic,ornelas2024non,psaroudaki2023skyrmion,kam2026quantum}. In contrast, our constructed entangled state is generated from two complex quantum states, each of which jointly contains both polarization and a spatial-mode distribution (see Eq.~\ref{u_dis}), where both single-photon reduced states $\hat{\rho}_A = \mathrm{Tr}_B\ket{\Psi}_{A,B}\bra{\Psi}_{A,B}$ and $\hat{\rho}_B =\mathrm{Tr}_A\ket{\Psi}_{A,B}\bra{\Psi}_{A,B}$ are by design a skyrmion in the sense of Eq.~\ref{eq:denisty matrix}. Interestingly, for $\hat{\rho}_B$ the eigenvectors are replaced by their complex conjugates, resulting in a skyrmion of opposite type to that of $\hat{\rho}_A$, as can be seen in Fig. \ref{fig:2}(b).

\begin{figure*}[ht]
    \centering
    \includegraphics[width=\textwidth]{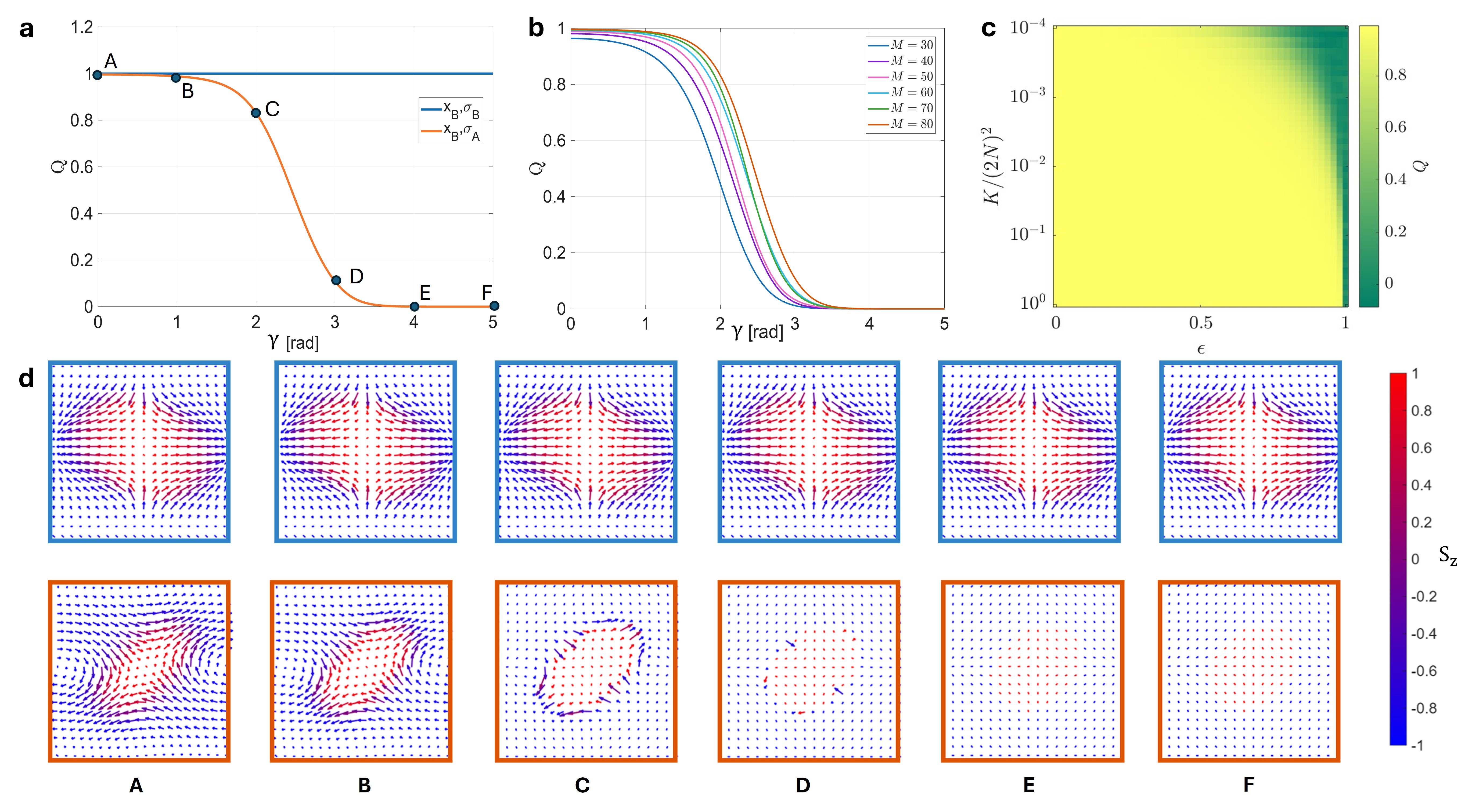}
        \caption{\textbf{Robustness of local and non-local skyrmions.} \textbf{(a)} Skyrmion number $Q$ in local and nonlocal reduced subspaces ($M=80$) as a function of the dephasing strength $\gamma$ for a Gaussian phase-noise channel applied to the two-photon density matrix (Eq. \ref{eq:matrix}). The skyrmion number of the local reduced subspace $x_B,\sigma_B$ (blue) remains invariant under two-photon dephasing, while the non-local reduced subspace ($x_B,\sigma_A$) (orange) remains invariant up to $\gamma\sim 1.5$, indicating the gradual destruction of the non-local skyrmion. The textures in panel (\textbf{d}) visualize this transition. \textbf{(b)} Topological charge $Q$ for a non-local skyrmion as a function of the dephasing strength $\gamma$ for a Gaussian phase-noise channel and for different number of modes $M$. The figure illustrates that the resilience of the non-local skyrmion to noise increases with the number of modes.  \textbf{(c)} Robustness of the non-local skyrmion to additive noise. We mix the two-photon density matrix ($M=60$) with a random Wishart density matrix~\cite{zyczkowski2011generating,collins2016random} of rank $K$, with mixing probability $\epsilon$. The skyrmion number remains invariant under a wide range of noise ensembles and mixing probabilities.
        \textbf{(d)} As the dephasing strength $\gamma$ increases in the Gaussian phase-noise channel, the skyrmion undergoes a transition from a Néel-type texture to a Bubble-type texture, as illustrated by the different stages shown in panel~(a)~\cite{nagaosa2013topological,mangold2026phonon,fullerton2026observation}. See Supplementary Material Section~S4.4 for the skyrmion densities corresponding to the points (A)-(F) in panel 3(d), and for the noise analysis for different numbers of modes corresponding to panel 3(b).}
    \label{fig:3}
\end{figure*} 

Remarkably, we numerically observe a new type of non-local skyrmion, in the reduced state that emerges within the hybrid subspace spanned by photon $A$’s polarization and photon $B$’s spatial mode, or vice versa ($\hat{\rho}_{A_\sigma B_x}=\mathrm{Tr}_{\sigma_B,x_A}\ket{\Psi}_{A,B}\bra{\Psi}_{A,B}$ and $\hat{\rho}_{A_x B_\sigma}=\mathrm{Tr}_{\sigma_A,x_B}\ket{\Psi}_{A,B}\bra{\Psi}_{A,B}$, respectively), as can be seen in Fig. \ref{fig:2}(c). These reduced non-local skyrmions are of the Bloch type \cite{liu2016skyrmions} whereas the single-photon reduced states and the two-photon state form Néel type skyrmions. 

\textbf{\textit{Nesting of topological charge.}} The hierarchy of coexisting topological textures motivates a formal definition of what we term nested topology. In contrast to a conventional skyrmion, which is identified by a single state representation, here the topological content of the two-photon state is distributed across multiple reduced representations obtained by partial tracing over complementary degrees of freedom.
We say that the topology of $\Psi_{A,B}$ is nested if nontrivial skyrmion charges persist in more than one reduced representation.
In this sense, the two-photon skyrmion decomposes into a structured collection of local and hybrid, non-local topological quasiparticles, whose topological charge is derived directly from the full two-photon wavefunction, as evidenced from Fig. \ref{fig:2}. By contrast, previous realizations of quantum optical skyrmions ~\cite{ma2025nanophotonic,ornelas2024non,psaroudaki2023skyrmion,kam2026quantum} are not nested, as the skyrmion texture only exists in a specific two-photon state but not in a reduced single-photon subsystem.

Beyond classification, nested topology provides a practical advantage for robustly encoding information on the topology of a density matrix. Since the topological content of the two-photon state directly influences multiple reduced representations, access to only a subset of system degrees of freedom does not erase the topological signature. In particular, even under severe loss or decoherence, nontrivial quantized topological charges may remain observable in a local or non-local manner. 
To show this, we model the evolution of a two-photon entangled skyrmion under a stochastic dephasing channel by introducing a Gaussian-distributed random phase acting on the coherence terms of the two-photon density matrix: 
\begin{equation}
\begin{aligned}
    \hat{\rho}_{A,B}\mapsto\frac{1}{2}\ket{u_{1}u_{1}^*}\bra{u_{1}u_{1}^*}
    +\frac{1}{2}\ket{u_{2}u_{2}^*}\bra{u_{2}u_{2}^*}\\+\frac{1}{2}\exp(i\hat{\phi})\ket{u_{1}u_{1}^*}\bra{u_{2}u_{2}^*}\\+\frac{1}{2}\exp(-i\hat{\phi})\ket{u_{2}u_{2}^*}\bra{u_{1}u_{1}^*}.
\end{aligned}
\label{eq:matrix}
\end{equation}
The ensemble average suppresses the off-diagonal elements as 
$\langle\exp(i\hat{\phi})\rangle=\exp(i\mu-\frac{\gamma^2}{2})$, thereby reducing the inter-photon coherence supporting a non-local skyrmion texture. This channel can manifest as statistically independent random phase noises acting on either or both modes $u_1,u_2$ and for either or both photons.
Note that Eq.~\ref{eq:matrix} introduces a deliberately selected non-local dephasing channel. This channel is not intended to represent a completely general noise model, but rather to demonstrate how nested topology can preserve topological information across different reduced representations under a decoherence process. Fig. \ref{fig:3}(a) shows that as the phase uncertainty increases, both the local and the non-local skyrmions remain robust to the dephasing noise. Beyond a certain level of noise, which depends on the number of modes used for encoding (as shown in Fig. \ref{fig:3}(b)), the system undergoes a topological transition from a regime where local and non-local skyrmions coexist to a regime in which entanglement is lost, yet the local skyrmion encoded on each photon separately survives over the noisy channel~\cite{ornelas2025topological}. In principle, for an asymptotically large number of encoding modes, our non-local skyrmion - as opposed to other quantum skyrmions - is also immune to this noise, owing to the fact that the noise shifts the skyrmion distribution in the density matrix from Bloch-type to Bubble type rather than completely destroying its topology ~\cite{nagaosa2013topological,mangold2026phonon,fullerton2026observation} (Fig. \ref{fig:3}d). In a bubble-type skyrmion, the texture is dominated by the out-of-plane component, which reverses direction across a narrow domain-wall region. The resulting localization of the skyrmion density makes the topological charge more difficult to resolve using the same number of encoding modes. See Supplementary Material, Sections~S4.4, for the corresponding noise analysis, where the skyrmion-density distributions are also presented.

\begin{figure*}[ht]
    \centering
    \includegraphics[width=1.7\columnwidth]{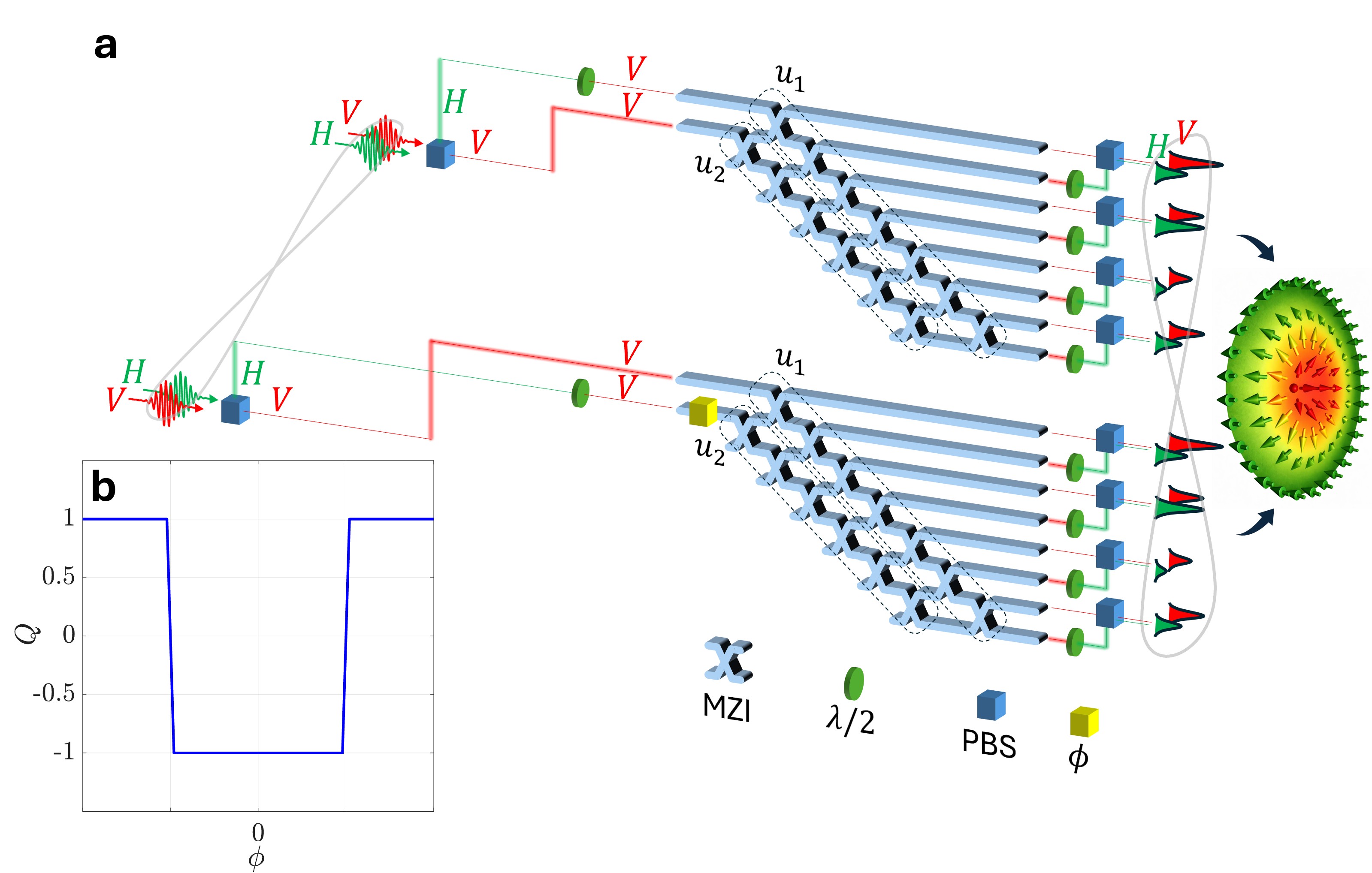}
        \caption{\textbf{Experimental concept for generating and tuning a two-photon skyrmionic state on chip.} \textbf{(a)} A pair of entangled photons is prepared in a Bell state encoded in two paths and two polarizations. The two polarizations in each path are further separated into two spatial bins by a polarizing beam splitter (PBS) and converted to the same $V$ polarization by a half-wavelength plate. Each of the paths is then incident upon a reconfigurable, integrated interferometric network which implements a controlled mapping from two incident spatial bins into two designed eigenmodes $\ket{u_1}$ and $\ket{u_2}$. At the output, every other spatial bin is converted back into the $H$ polarization, and recombined with its neighboring output bin on a PBS. This realizes an entangled skyrmion in the desired eigenmode basis. \textbf{(b)} Skyrmion number $Q$ extracted from the measured Stokes texture as a function of the programmable relative phase $\phi$. The topological charge exhibits a sharp transition between opposite charges, demonstrating that the skyrmion topology can be used for enhanced sensing of phase objects.}
    \label{fig:4}
\end{figure*}

Fig.~\ref{fig:3}(c) shows the robustness of the nested skyrmionic topology to additive noise. We consider the following convex mixture of the two-photon density matrix $\hat{\rho}_{A,B}=(1-\epsilon)\ket{\Psi}_{A,B}\bra{\Psi}_{A,B} +\epsilon\hat{\sigma}_{A,B}$ where $\epsilon\in[0,1]$ determines the relative weight of the additive noise. The random state $\hat{\sigma}_{A,B}$ is a $(2M)^2\times(2M)^2$ random Wishart density matrix with $K$ degrees of freedom~\cite{zyczkowski2011generating,collins2016random} generated according to
\begin{equation}
\hat{\sigma}_{A,B}
=
\frac{GG^{\dagger}}
{\operatorname{Tr}\!\left(GG^{\dagger}\right)},
\qquad
G\in\mathbb{C}^{(2M)^2\times K},
\end{equation}
where the entries of $G$ are independent complex Gaussian random variables. The parameter $K\leq (2M)^2$ determines the rank of $\hat{\sigma}_{A,B}$. In essence, this construction tests our skyrmions' resilience against all types of additive noise (amplitude, phase and polarization).
We plot the skyrmion number of the non-local system $\hat{\rho}_{x_A\sigma_B }$ as a function of $K$ and $\epsilon$, displaying the robustness of the skyrmion number to different noise ensembles. As $K$ increases, the noise in the reduced subsystem approaches a maximally mixed state, thus mimicking a depolarization channel, whereas the skyrmion number becomes more robust. Indeed, it can be readily shown from Eq.~\ref{eq2} that the skyrmion texture is invariant under the action of a depolarization channel, in agreement with our numerical observation.   

\textbf{\textit{Nested topology of multi-photon states.}} Intriguingly, we find that the bipartite nested topology can manifest in multi-photon mixed states. Namely, we find a family of $N$-photon density matrices $\hat{\rho}_{s_1s_2...s_N}$ that, upon tracing out any $N-2$ photons, leaves the system in a bipartite state $\hat{\rho}_{s_i s_j}=\mathrm{Tr}_{\lbrace s_k|k\neq i,j\rbrace}\hat{\rho}_{s_1 s_2...s_N}$ with a nested topology. That is, there exists a topologically nontrivial skyrmion texture in both the local and non-local subspaces of $\hat{\rho}_{s_i s_j}$, for any photon pair $i,j$.   

The construction is as follows: first, we note that the Bell state $\ket{\Psi_{\textrm{Bell}}} = \left(\ket{u_1}_A\ket{u_1}_B+e^{i\phi}\ket{u_2}_A\ket{u_2}_B\right)/\sqrt{2}$ (that is, a state where the mode $u_i$ of $B$ is not conjugated), is also a nested topological state, as presented in the Supplementary Material, section S3. We find numerically that a nested topology multiphoton state is constructed from the Bell state as the following biseparable mixed state
\begin{equation}
    \hat{\rho}_{s_1s_2...s_N} = \frac{1}{\binom{N}{2}} \sum_{i<j}\ket{\Psi_{\mathrm{Bell}}}_{s_i s_j}\bra{\Psi_{\mathrm{Bell}}}_{s_i s_j}\bigotimes_{k\neq i,j}\hat{\varrho}_k 
\label{eq:rhos1n}
\end{equation}
where $\hat{\varrho}$ is a single-photon density matrix, to be specified forthwith. Eq.~\ref{eq:rhos1n} describes a convex mixture of states where systems $s_i,s_j$ are prepared in the Bell state where the rest are prepared in a tensor product of identical single photon states. The reduced bipartite state for any $i,j$ becomes
\begin{equation}
\begin{split}
\hat{\rho}_{s_is_j}&=\frac{2}{N(N-1)}\ket{\Psi_{\mathrm{Bell}}}_{s_i s_j}\bra{\Psi_{\mathrm{Bell}}}_{s_i s_j} \\&+\frac{2(N-2)}{N(N-1)}[\hat{\rho}_{i}^{(0)}\otimes \hat{\varrho}_j+\hat{\varrho}_{i}\otimes \hat{\rho}^{(0)}_j] \\&+\frac{(N-2)(N-3)}{N(N-1)}\hat{\varrho}_i\otimes \hat{\varrho}_j
\end{split}
\end{equation}
where $\hat{\rho}_{i}^{(0)}=\mathrm{Tr}_{s_j} \ket{\Psi_{\mathrm{Bell}}}_{s_i s_j}\bra{\Psi_{\mathrm{Bell}}}_{s_i s_j} =\left(\ket{u_1}\bra{u_1} + \ket{u_2}\bra{u_2}\right)/2$ is the single-photon density matrix of Eq. \ref{eq:denisty matrix}. The resulting state is a mixture between the pure two-photon skyrmion state $\ket{\Psi_{\mathrm{Bell}}}$ and two noise terms that depend on $\hat{\varrho}$. As a result, different choices of the matrix $\hat{\varrho}$ may, in general, hinder the structure of the resulting nested skyrmion textures. In the Supplementary Material, section S3 we show that a choice of a diagonal and sparse $\hat \varrho$ (that is, an incoherent mixed state where only a part of the modes are occupied) leads to a minimal interference with the nested skyrmion textures encoded in the two-photon Bell state $\ket{\Psi_{\mathrm{Bell}}}$. 

\textbf{\textit{Proposed experimental implementation.}} Figure~\ref{fig:4} illustrates the proposed experimental implementation of the two-photon state used in this work. In Fig.~\ref{fig:4}(a), a polarization-entangled photon pair is prepared in a Bell state of the form $\ket{\Psi_{\mathrm{in}}}_{A,B}=\frac{1}{\sqrt{2}}\left(\ket{H}_{A}\ket{V}_{B}+\ket{V}_{A}\ket{H}_{B}\right)$, where $A$ and $B$ denote photons in different paths. This state serves as the entanglement resource for the subsequent mode engineering stage. 

Each path then passes through a polarizing beam splitter (PBS) that separates the $H$ and $V$ modes, and the horizontal polarization goes through a $\lambda$/2 plate, thus mapping the two polarization states into two spatial bins. Next, the beams are coupled to a photonic circuit that converts the two input ports into a pair of engineered modes \(\{\ket{u_1},\ket{u_2}\}\). This mapping is implemented using a programmable mesh of Mach--Zehnder interferometers (MZIs) \cite{roques2024measuring,roques2025automated} and a controllable phase shifter \(\phi\) at one of the inputs. Operationally, the circuit performs a coherent basis transformation from polarization qubit states to the target modal basis, while preserving the bipartite entanglement inherited from the input state. At the output, every other spatial bin is converted back to the $H$ polarization and recombined with its neighboring bin using a PBS. In this way, the two-photon output state is synthesized as $\ket{\Psi_{\mathrm{out}}}_{A,B}=\frac{1}{\sqrt{2}}\left(\ket{u_1}_A\ket{u_1}_B+e^{i\phi}\ket{u_2}_A\ket{u_2}_B\right)$. In practice, the MZI meshes provide a scalable and programmable platform to realize these mode transformations with high fidelity~\cite{miller2013self, miller2020analyzing,bogaerts2020programmable,mor2026separating,miller2025universal}, and naturally support extension to larger sets of eigenmodes (higher Schmidt rank), by increasing the number of layers in the mesh (as discussed in Refs. \cite{roques2024measuring,roques2025automated}, each layer $i=1,...,d$ in the mesh shapes an eigenmode $\ket{u_i}$ in the bipartite state $\ket{\Psi}=\sum_{i=1}^d \lambda_i \ket{u_i}_A\ket{u_i}_B$. Thus, cascading more than two layers allows for higher Schmidt ranks $d>2$). While the proposed mesh is implemented for spatial modes, it can readily be implemented for spectral bins as well~\cite{karnieli2025variational}.

Figure~\ref{fig:4}(b) shows the designed topological charge \(Q\) as a function of the relative phase \(\phi\), displaying a highly sensitive, phase-controlled switching between topological sectors. This numerically predicted phase sensitivity provides a direct experimental knob for enhanced sensing of phase objects using topological quasiparticles encoded in density matrices.

To measure the skyrmionic reduced density matrices, one would require a full quantum state tomography of all degrees of freedom, which in general would scale quadratically with the number of modes $O(M^2)$. However, since the density matrices are sparse with rank $d$, the number of measurements should scale as $O(Md)$ (see Supplementary Material, Sections~S4.6 for Measurement and preparation resource scaling). We note that instead of full quantum state tomography, one could use a partially coherent light analyzer \cite{roques2024measuring,mor2026separating} to measure the density matrix. In that case, a designated photonic circuit learns the eigenstate decomposition of an input density matrix $\rho=\sum_i p_i \ket{u_i}\bra{u_i}$ in descending eigenvalue order.

\section*{Discussion and Conclusions}
Our results establish that skyrmionic topology can be encoded and manipulated directly at the level of the density matrix of mixed quantum states, rather than requiring a fully pure state, which is often fragile. The key step is to treat the density matrix as an effective field over a two-dimensional coordinate space, and to define a normalized coherence-Stokes texture $\mathbf{s}(x,x')$ whose winding supports an integer skyrmion charge. This viewpoint unifies classical partial coherence and single-photon density matrices, and provides a natural bridge between classical field topology and quantum state structure.

A central consequence of this framework is a reduction in the dimensional resources required to encode topological quasiparticles. While traditional encoding schemes require a two dimensional distribution, encoding skyrmions in density matrices requires only one dimension. We further find that our skyrmion encoding admits a sparse realization: a low-rank density matrix is sufficient to reproduce the designed charge even as $M$ is increased to improve the resolution of the skyrmion texture, and numerically the minimal rank follows $d=|Q|+1$. Practically, this compresses the sampling requirements and reduces the number of interferometric degrees of freedom needed for preparation and measurement, suggesting a scalable route to multiplexing multiple topological channels within a limited number of modes.

Extending this framework to photon pairs reveals a phenomenon we term nested topology. Unlike conventional approaches, where topology is encoded solely in a two-photon wavefunction or in a single-photon wavefunction, an entangled skyrmion in our setting generates multiple interdependent topological charges across reduced subsystems: the local state of each photon, as well as hybrid reductions combining one photon’s internal degree of freedom (pseudospin) with the other's external degree of freedom (mode space). This hierarchy of topological invariants provides a structured way to interrogate the distribution of correlations across degrees of freedom, and suggests that information about the topology of the system can persist in experimentally accessible subspaces even when the full quantum state of the system is not reconstructed.

The nested topology in the bipartite system also affords robustness under environmental noise. Remarkably, we observe that such skyrmions are particularly resilient against strong additive noise or the suppression of inter-photon coherence. After a certain point, which depends on the number of modes in the encoding, the topological charge of the non-local skyrmions vanishes, while the local skyrmions remain unchanged over the entire noise range. This yields a noise-driven transition from a fully nested regime to a locally nested regime, in which topology based on entanglement is lost, yet the information about said topology remains locally encoded on each photon. We then show that the concept of nested topology further extends to multi-photon mixed states, where tracing-out all but two photons in the state leaves the reduced subsystem in a nested topological quantum state. 

Looking ahead, the present framework may be extended to other classes of topological quasiparticles such as merons~\cite{krol2021observation,ghosh2021topological} and hopfions~\cite{shen2023topological,wan2022scalar,wu2022hopfions}, which are promising candidates for future density-matrix encodings. Such generalizations may reveal different forms of nested topology and distinct robustness characteristics. However, their explicit construction and characterization lie beyond the scope of the present work.

Furthermore, our density matrix description is not limited to quantum optics, and can extend to any few- or many-body quantum systems fulfilling the correct requirements, regardless of the properties of the underlying particles. For example, our formalism can be implemented in arrays of trapped ions or neutral atoms using two long-lived energy states (either ground or metastable) as the pseudospin; and the sideband spectrum created by trap phonons~\cite{shaw2025erasure,home2009complete} or the manifold of Zeeman sub-levels as a synthetic dimension. 

Finally, we proposed a practical quantum optical implementation based on integrated photonics that provides a concrete experimental route towards realization of two-photon entangled skyrmions. The same platform naturally supports extensions to larger mode sets (higher $|Q|$ and higher-rank constructions), to alternative degrees of freedom (time-bin or frequency), and for enhanced sensing of phase objects due to the sharp topological transitions in partially coherent quantum states.

Note: During the preparation of this manuscript, we became aware of related works that considered skyrmions in partially coherent light\cite{liu2026incoherent, chen2026imprinting, martinez2026effective}. While our formalism can also describe partially coherent light, the present work goes beyond this regime by developing a fully quantum framework in which skyrmionic topology is encoded directly in the density matrix of light. Moreover, our framework extends to multipartite quantum states, enabling skyrmion wavefunctions with nested topological structures.

\newpage
\section*{Acknowledgements}

The authors thank Guy Bartal and Meir Orenstein for fruitful discussions. Am.K. acknowledges the support from the Azrieli fellowship and the support from the Helen Diller Quantum Center at the Technion. S.T. acknowledges support from the Rothschild fellowship of the Yad Hanadiv foundation, the VATAT Quantum fellowship of the Israel Council for Higher Education, the Helen Diller Quantum Center postdoctoral fellowship and the Viterbi fellowship of the Technion - Israel Institute of Technology. C.R-C is supported by a Stanford Science Fellowship and acknowledges startup funding from the Institute of Science and Technology, Austria (ISTA). Av.K. acknowledges support of the VATAT Quantum fellowship for new faculty, by the Israel Council for Higher Education; and the Deloro Career Advancement Chair in Engineering by the Technion - Israel Institute of Technology.

\section*{Supporting Information}
Supplemental material is available free of charge and includes the following sections:
\begin{itemize}
    \item[\textbf{S1}] Basic definitions. 
    \item[\textbf{S2}] Realization of higher-charge and distinct skyrmion textures, including the construction of skyrmionium textures in coherence space.
    \item[\textbf{S3}] Nested topology of quantum skyrmions encoded in Bell states.
    \item[\textbf{S4}] Calculation of the skyrmion number using the differential, Berg-Lüscher, and McWilliam et al. methods, together with analyses of convergence with the number of modes, robustness to noise, and measurement and preparation resource scaling.
\end{itemize}

\section*{Abbreviations}

\begin{itemize}
    \item[1D] One-dimensional.
    \item[MZI] Mach-Zehnder interferometer.
    \item[PBS] Polarizing beam splitter.
    \item[PSD] Positive semidefinite.
\end{itemize}

\newpage
\section*{Declarations}
\begin{itemize}
\item \textbf{Funding:}
This research received no external funding.
\item \textbf{Competing interests:}
The authors declare no competing interests.
\item \textbf{Ethics approval and consent to participate:}
This manuscript did not relate to any human or animal experiment
\item \textbf{Data availability:}
The data that supports the findings of this study is available from the corresponding author upon reasonable request.
\item \textbf{Code availability:} 
The code used in this study is available from the corresponding author upon reasonable request.
\item \textbf{Author contribution:} Av.K., Am.K. and S.T. conceived the project. Av. K. supervised the research. Am.K., Av. K. S.T. and C.R-C wrote the manuscript.
\end{itemize}

\printbibliography[prenote=referencecount]

\newpage

\section*{For Table of Contents Use Only}

\textbf{TABLE OF CONTENTS GRAPHIC}
\begin{itemize}
    \item \textbf{Manuscript Title:} ''Quantum Skyrmions in Mixed States of Light and their Nested Topology".
    \item \textbf{Authors:} Amit Kam, Charles Roques-Carmes, Shai Tsesses, and Aviv Karnieli.
    \item \textbf{Synopsis:} The graphic illustrates the encoding of skyrmionic topological textures within the density matrix of mixed states of light. The coherence-Stokes vector maps correlations between the photonic pseudospin and external optical modes onto a skyrmion texture. In bipartite systems, this topology can be nested across multiple reduced pseudospin–mode subspaces, revealing simultaneous local and nonlocal topological structures.
\end{itemize}

\rule{0.05in}{1.75in}%
\begin{minipage}[b][1.75in][c]{3.25in}
    \sffamily
    \frenchspacing
    \centering
    \includegraphics[
        width=3.25in,
        height=1.75in,
        keepaspectratio
    ]{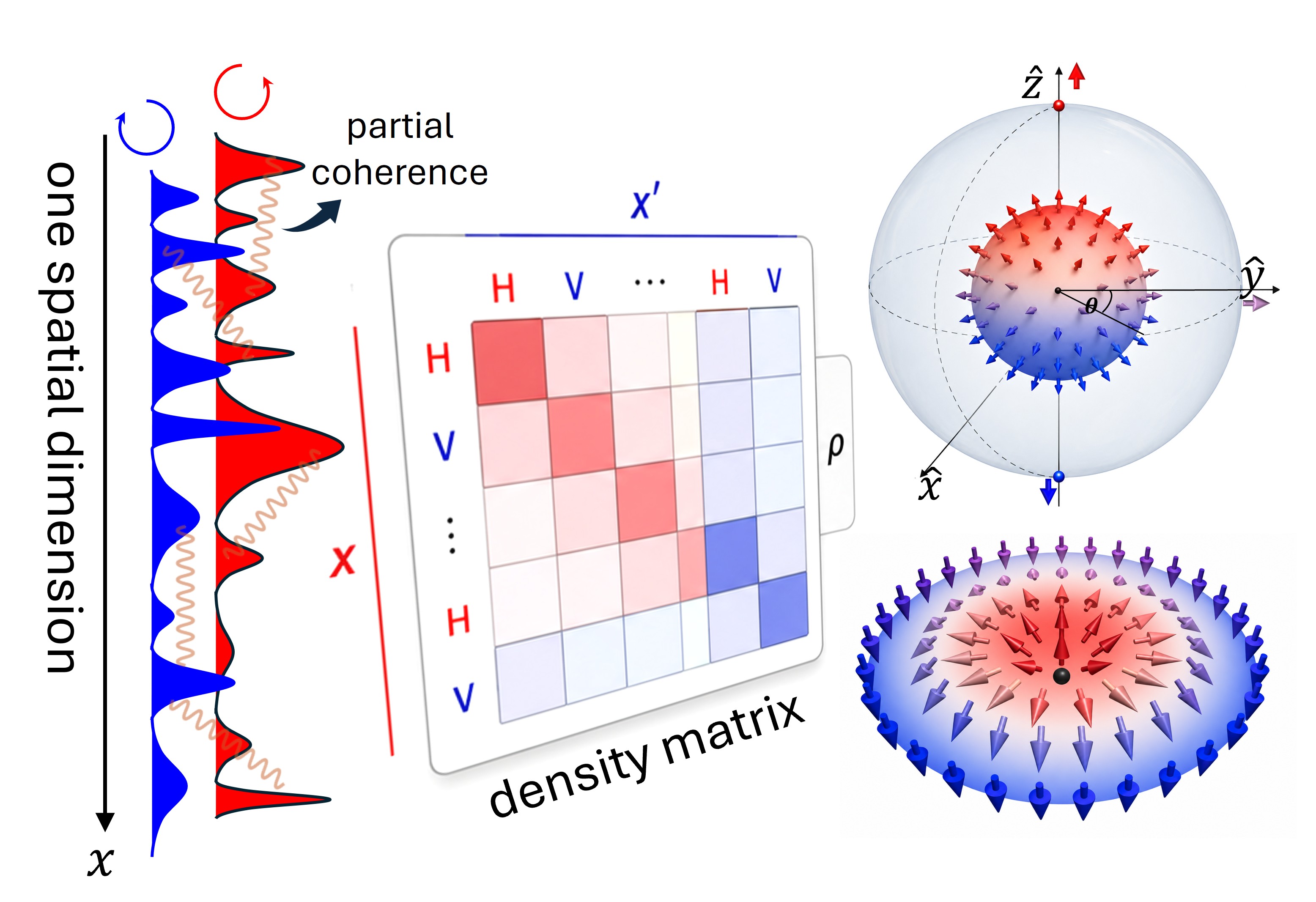}
\end{minipage}%
\rule{0.05in}{1.75in}

\end{document}